\documentclass[12pt,draftclsnofoot,onecolumn]{IEEEtran}
\usepackage{setspace}
\usepackage{clrscode}
\usepackage{amsmath}
\usepackage{amssymb}
\usepackage[dvips]{graphicx}
\usepackage{epsfig}
\usepackage{hyperref}
\usepackage{bm}
\usepackage{subfigure}
\usepackage{caption}

\newcommand{\y}{\mathbf{y}}
\newcommand{\h}{\mathbf{h}}
\newcommand{\B}{\mathbf{B}}
\newcommand{\z}{\mathbf{z}}
\newcommand{\w}{\mathbf{w}}
\newcommand{\U}{\mathbf{U}}
\newcommand{\dd}{\mathbf{s}}
\newcommand{\bb}{\mathbf{b}}
\newcommand{\figsize}{0.65}
\newcommand{\C}{\mathbb{C}}
\newcommand{\PP}{\mathbf P}
\newcommand{\A}{\mathbf{A}}
\newcommand{\HH}{\mathbf{H}}
\newcommand{\T}{\mathbf{T}}
\newcommand{\0}{\mathbf{0}}

\newcommand{\tout}{\text{\footnotesize{out}}}
\newcommand{\sout}{\text{\footnotesize{cell,out}}}

\newcommand{\E}{\mathbb{E}}
\newcommand{\ssnr}{\text{\scriptsize{SINR}}}
\newcommand{\snr}{{\text{\footnotesize{SINR}}}}

\newtheorem{Lem1}{Proposition}
\newtheorem{Lem}{Theorem}
\newtheorem{Rem}{Remark}
\newtheorem{Exm}{Example}
\newtheorem{Corr}{Corollary}

\begin{document}

\title{Outage Analysis of Relay-Assisted mmWave Cellular Systems Employing JSDM}
    \author{
\IEEEauthorblockN{Jun Chen and Deli Qiao}

\thanks{The authors are with the School of Information Science and Technology, East China Normal University, Shanghai, China 200241. D. Qiao is also with Shanghai Key Laboratory of Multidimensional Information Processing, East China Normal University, Shanghai, China 200241. Email: 51161214011@stu.ecnu.edu.cn, dlqiao@ce.ecnu.edu.cn.}
\thanks{This work has been supported in part by the National Natural Science Foundation of China (61671205) and the Shanghai Sailing Program (16YF1402600). }
\thanks{This paper will be presented in part at the IEEE International Conference on Communications (ICC), Kansas City, MO, 2018 \cite{icc18}. }}
\maketitle

\begin{abstract}
In this paper, the outage performance of relay-assisted millimeter wave (mmWave) cellular systems employing joint spatial division and multiplexing (JSDM) is investigated. It is assumed that the macro base station (BS) equipped with a large number of antennas serves the single antenna pico BS (as a relay) and users simultaneously, and that the pico BS is located at the edge of the macro cell. Theoretical analysis of the signal-to-interference-plus-noise ratio (SINR) outage probability of each user is first obtained. The cell SINR outage probability is then derived. Under the noise-limited assumption, simplified closed-form expressions of the outage probability are given as well. Simulation results demonstrating the performance improvement due to the relay introduced by the pico BS are provided. Overall, the impact of deploying pico BS as a relay in the mmWave cellular systems is characterized.
\end{abstract}

\section{Introduction}
With the vast use of smart phones, tablets and social networks, the demand of high data rates has surged recently. The fifth generation (5G) cellular system has been put forward to address the exponentially increased demands in mobile data traffic. In this regards, different techniques have been proposed by the academia and the industry, e.g., massive multiple input multiple output (MIMO) \cite{tom-noncooperative}, \cite{massiveMIMO}, millimeter wave (mmWave) \cite{tsrapp}, \cite{2},  and ultra-dense networks  \cite{1}, \cite{udn}. Of particular interest is the mmWave communication systems \cite{mmwavetutorial}. Because of the vast amount of spectrum in the underutilized mmWave frequency bands, mmWave communication systems can offer an order of magnitude increase in achievable rate compared with current cellular systems and play an important role in future cellular networks \cite{1}.

Despite the great potential of high data rates with mmWave systems due to the richness in bandwidth, the transmission range is generally limited due to the high free-space path loss and poor penetration in high frequency \cite{tsrapp}. Thus, it is highly possible that some users may fall in outage in mmWave systems, and hence the coverage analysis of mmWave systems have attracted much attention recently (see e.g., \cite{coverage}-\cite{BScooperation} and references therein). For instance, the authors have proposed a stochastic geometry framework for analyzing the coverage and rate of mmWave systems assuming pencil beams for the users in \cite{coverage}. They have shown that there is an optimal relative base station (BS) density for the signal-to-interference-plus-noise ratio (SINR) and rate performance beyond which the performance doesnot improve in a dense mmWave network. An analytical framework which computes coverage probabilities and rate of mmWave cellular networks has been proposed in \cite{stommWave}, where path-loss and blockage models based on empirical data for mmWave propagation have been taken into account. Relying on the noise-limited assumption for modeling mmWave cellular systems, simple and closed-form formulas for computing the coverage probability and the average rate have been obtained. In \cite{gursoycoverage}, the authors have made a further step and investigated the coverage in heterogeneous mmWave networks with homogeneous Poisson point process (PPP) models of the BS and user distributions, where beamforming with pencil beam at the BSs is assumed. It has been shown that biasing towards the small cells in user association can improve both the coverage probability and the rate. Also, the authors have assumed homogeneous PPP model for the macro cells while Poisson hole process (PHP) model for small cells and studied the coverage of the proposed non-uniform mmWave heterogeneous cellular network in \cite{wcsp}. They have also shown that there exists an optimal density of the small cells to achieve the best coverage probability.  The performance of relay-assisted mmWave systems has also been characterized in \cite{mmwaverelay}. Note however that the above works generally assume ideal pencil beams for beamforming. In practical mmWave systems, large antenna arrays, or massive MIMO, are usually deployed at the macro BSs, in which case the ideal sector beam may not hold and different approximate beam patterns have been incorporated for coverage analysis in \cite{mmwaveantenna}. The authors have used stochastic geometry tools to carry out a comprehensive investigation on the impact of directional antenna arrays in mmWave networks in \cite{mmwaveantenna}. In \cite{BScooperation}, the authors have considered the problem of BS cooperation in mmWave heterogenous network and shown that BS cooperation through jointly beam steering can increase the coverage probability for a typical user in some cases.

Moreover, for mmWave systems employing massive MIMO, a two-stage precoding scheme, joint spatial division and multiplexing (JSDM) has been proposed in \cite{jsdmmmwave}. The idea of JSDM is to make use of the channel covariance information to reduce the channel estimation overhead and mitigate the interference for users in different groups partitioned according to channel covariance subspaces \cite{jsdm}. The authors have shown that taking advantage of the highly directional channel characteristics, JSDM can achieve remarkable sum rate and simplify system operations \cite{jsdmcoverage}. Nevertheless, the coverage analysis associated with the mmWave communication systems employing JSDM is still lacking.

%

In this paper, we investigate the outage performance of a relay-assisted mmWave cellular system employing JSDM, where a macro BS serves the users with the aid of a pico BS (as a relay) and the users are independently and uniformly distributed in the cell. We assume that the macro BS employs JSDM to serve the users and the pico BS simultaneously, and the pico BS works in full-duplex mode and employs decode-and-forward (DF) to forward the data from the macro BS to its served users. We note that the authors in \cite{dfbrs} have also investigated the mmWave networks with DF relays and shown the coverage improvement due to the DF relays, albeit ideal pencil beams have been assumed. In this work, we consider a two-tier mmWave cellular system, where the pico BS serves as a relay. We first consider a specific user grouping and then the random groups due to the random user locations. The main contributions of this work are summarized as follows.

\begin{itemize}
  \item  We propose a general analytical framework to analyze the outage performance in mmWave cellular networks employing JSDM, where a new cell association strategy based on the relay channel is incorporated.

  \item  We derive the theoretical expressions of the average user and cell outage probabilities.

  \item Numerical results in accordance with the theoretical analysis are provided as well. Through numerical results, we demonstrate that mmWave systems employing JSDM is still noise-limited and employing pico-BS as a relay can improve the coverage probabilities.
\end{itemize}

The paper is organized as follows. The system model and the preliminaries on the user association and JSDM are briefly introduced in Section II. Section III discusses the outage analysis of the relay-assisted mmWave systems in detail. Numerical results are provided in Section IV. Finally, Section V concludes this paper, with some lengthy proofs in the Appendix.

\section{System Model and Preliminaries}

\subsection{System model}
\begin{figure}
    \centering
    \includegraphics[width=0.45\textwidth]{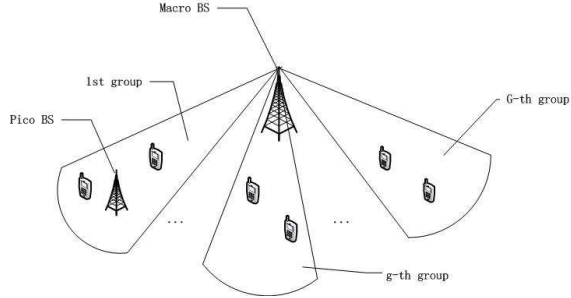}
    \caption{System model.}
    \label{fig:cellmodel}
\end{figure}

As shown in Fig. \ref{fig:cellmodel}, we consider the single-cell scenario in which a macro BS equipped with $M$ antennas with uniform linear array (ULA) serves $K$ single-antenna users with the aid of a single-antenna pico BS located at the edge of the cell \footnote{ Note that since we employ JSDM at the macro BS in this work, the pico BS is assumed to lie in one group of the users partitioned by the channel covariance subspace as will be detailed later.}. Both the macro and pico BSs work in the same frequency band. It is assumed that the users are independently and uniformly distributed. Depending on the design of the cell association, certain users will be served by the full-duplex pico BS (as a relay) within the coverage of the macro BS. Note that we assume that the pico BS employs DF to process and retransmit the message sent from the macro BS to the user in pico-cell. We assume that full-duplex can be achieved with perfect self-interference cancellation through analog and digital cancellation \cite{fd-2}, \cite{fd-3}. We further assume that the macro BS employs JSDM for data delivery to the users and the pico BS. Denote the radius of the macro-cell as $R$ and the radius of the pico-cell as $r$. Let $P_m$ and $P_s$ be the transmission power levels at the macro and pico BS, respectively.


\subsection{Association Strategy}\label{sec:association}

Typically, the user is served by the BS with the smallest path-loss, i.e., the user is associated with the macro BS if
\begin{align}\label{eq:asso1}
\kappa^2P_md_{mu}^{-\alpha} \ge \kappa^2P_sd_{su}^{-\alpha},
\end{align}
where $d_{mu}$ and $d_{su}$ denote distance between the user and the macro and pico BS, respectively, $\alpha$ is the path loss exponent, and $\kappa^2=(\frac{\lambda_c}{4\pi})^2$ with $\lambda_c$ being the carrier wavelength \cite{coveragandrate}.

In this work, we consider a scenario that the pico BS serves as a relay. With the aid of the relay, it is possible to extend the coverage and improve the quality of communications of the mmWave systems. Note that the instantaneous rate of the relay channel with the DF protocol is given by the minimum rate of the two links \cite{tse}, i.e., the minimum rate of the link between the macro BS to the pico BS and the link between the pico BS and the user. Therefore, we consider the following cell association strategy such that the user is associated with the pico BS if
\begin{align}\label{eq:usera}
\min\{\kappa^2P_md_{ms}^{-\alpha},\kappa^2P_sd_{su}^{-\alpha}\}\geq \kappa^2P_md_{mu}^{-\alpha},
\end{align}
where $d_{ms}$ denotes the distance between the macro and pico BS. Denote $p_{gm}$ as the probability that the user in group $g$ is associated with the macro BS and $p_{gs}=1-p_{gm}$ as the probability that the user in group $g$ is associated with the pico BS. We will obtain the expressions for $p_{gm}$ and $p_{gs}$ in the following.

%

\subsection{Joint Spatial Division and Multiplexing with Per-Group Processing (JSDM-PGP)}

As shown in Fig. \ref{fig:cellmodel}, we consider the one-ring scattering model for the channel between the users, including the pico BS, and the macro BS. Taking into account the small scale fading only, the channel covariance matrix for a user in group $g$ with angle-of-arrial (AOA) $\theta_g$ and angular spread (AS) $\Delta_g$ is given by \cite{jsdmmmwave}
\begin{align}
[\mathbf{R}_g]_{m,p}=\frac{1}{2\Delta_g}\int_{-\Delta_g+\theta_g}^{\Delta_g+\theta_g}e^{-j2\pi D(m-p)\sin(t)}\, dt,
\end{align}
where $D$ denotes the distance between the adjacent antenna elements of the macro BS in terms of carrier wavelength. Assume that the eigenvalue decomposition of $\mathbf{R}_g$ is given by $\mathbf{R}_g=\U_g\boldsymbol{\Lambda}_g\U_g^H$, where $\mathbf{U}_g$ is a tall unitary matrix of dimensions $M\times r_g$, $\boldsymbol{\Lambda}_g$ is a $r_g\times r_g$ diagonal semi-positive definite matrix, and $r_g$ denotes the rank of $\mathbf{R}_g$. Thus, the small-scale fading channel of user $k$ in group $g$ can be written without loss of generality as
\begin{align}\label{eq:channel}
\h_{gk}=\U_{g}\boldsymbol{\Lambda}_g^{1/2}\w_{gk},
\end{align}
where $\w_{gk}\sim\mathcal{CN}(\boldsymbol{0},\boldsymbol{I}_{r_g})$ is an i.i.d. Gaussian random vector.

In order to facilitate the analysis, we rewrite (\ref{eq:channel}) as
\begin{align}\label{eq:hsk}
\h_{gk}&=[\U_{g},\0_{M\times (M-r_g)}]\begin{bmatrix}
\boldsymbol{\Lambda}_g^{1/2}  & \mathbf{0}_{r_g\times(M-r_g)}  \\
\mathbf{0}_{(M-r_g)\times r_g}  &    \mathbf{0}_{(M-r_g)\times(M-r_g)}
\end{bmatrix}
\begin{bmatrix}
\w_{gk}        \\
\0_{(M-r_g)\times1}
\end{bmatrix}\nonumber\\
&=\tilde{\U}_{g}\tilde{\boldsymbol{\Lambda}}_g^{1/2}\tilde{\w}_{gk},
\end{align}
where $\tilde{\U}_{g}$ and $\tilde{\boldsymbol{\Lambda}}_g^{1/2}$ are $M\times M$ matrices, and $\tilde{\w}_{gk}$ is $M\times 1$ vector.

We assume that JSDM-PGP, a two-stage transmission scheme with dimension reduced channel state information, is employed at the macro BS. In this scheme, $K$ users are divided into $G$ groups with $K_{g}$ users each group and $K=\sum_{g=1}^GK_{g}$, and the received signal at the users in group $g$ is given by
\begin{align}
\y_g=\HH_g^H\B_g\PP_g\dd_g+\sum_{g'\neq g}\HH_g^H\B_{g'}\PP_{g'}\dd_{g'}+\z_g,
\end{align}
where $\B_g=[\bb_{g1},\ldots,\bb_{gB_g}]$ is the first-stage precoding matrix of dimension $M\times B_g$ to reduce the dimension of the channel and null the inter-group interference, and $\PP_g$ is the second stage precoding matrix of dimension $B_g\times S_g$. $S_g$ denotes the number of data streams in group $g$ and $S_g\le B_g\le r_g$. Denote $S=\sum_{g}S_g$ as the total number of data streams. $\HH_g=[\h_{g1},\ldots,\h_{gK_g}]$ is composed the instantaneous channel state information of the users in group $g$. $\dd_g \in \C^{K_g\times1}$ is the transmitted signal for the users in group $g$ and $\z_g \in \C^{K_g\times1}$ is additive white Gaussian noise at the users with i.i.d. entries of zero mean and unit variance.

Generally, $\B_g$ has been designed based on the long-term channel statistics to null the inter-group interference, i.e., $\HH_g\B_{g'}\approx \mathbf{0}$, for all $g_{'}\neq g$ \cite{jsdmmmwave}. $\PP_g$ is decided by the equivalent channel $\HH_g\B_g$ seen by the users in group $g$. For PGP, it is assumed that $\PP_g=\mathbf{I}_{r_g}$ such that different data streams are delivered along different beams determined in the first-stage such that the instantaneous channel feedback can be avoided.

\section{SINR Outage Probability}
In this section, we consider a single-cell scenario with one macro and one pico BS \footnote{We would like to note that the analysis method in this paper can be extended to the multi-cell scenario with multiple macro and pico BSs easily by taking into account the interference from the BSs working in the same frequency band. Details are omitted in this paper since they are trivial and the results are similar to the ones in this section.}. We investigate the SINR coverage probability of the considered system model. Note that if some users are associated with the pico BS, then the pico BS will be viewed as one user of the macro BS and we have a relay channel for such users. We assume that the small scale fading channel $\h_{ms}$ between the macro and the pico BS also takes the form of (\ref{eq:channel}) in a specific group, while assuming Rayleigh fading for the small scale fading channel between the single-antenna pico BS and the users. 

\subsection{Cell Association Probability}

Through simple geometry analysis, we can characterize the probability that the user in group $g$ is associated with the macro and pico BS, respectively. Without loss of generality, we assume that the pico BS lies at the edge of the macro-cell with AoA $\theta_g$. Note that if AOA is not $\theta_g$, the following results can be updated accordingly.
\begin{Lem1}\label{prop:p1}
In the relay-assisted mmWave cellular systems with a macro BS and a pico BS, the probability that a user is associated with the macro BS and the pico BS in group $g$ is given by
\begin{align}
p_{gm}&=1-p_{gs},\label{eq:p1s}\\
p_{gs}&=\frac{ r^2(\frac{\pi}{2}+\theta+\frac{1}{2}\sin(2\theta))-d_{ms}^2(2\theta-\frac{1}{2}\sin(4\theta))}{\Delta_gR^2},\label{eq:p2s}
\end{align}
respectively, where
\begin{align}\label{eq:theta}
\theta=\arcsin\left(\frac{r}{2d_{ms}}\right).
\end{align}
\end{Lem1}
\emph{Proof:} See Appendix \ref{app:p1} for details. $\hfill\square$

\subsection{Signal-to-Interference-Plus-Noise Ratio (SINR)}

Taking into account the path loss effect, the received signal of a user $k$ in group $g$ served by the macro BS can be expressed as
\begin{align}
y_{mk}&=\underbrace{\kappa d_{mk}^{-\alpha/2}\h_{mk}^{H}\bb_{gk}s_{gk}}_{\text{useful signal}}+\underbrace{\sum_{k^{'}\ne k}\kappa d_{mk}^{-\alpha/2}\h_{mk}^{H}\bb_{gk^{'}}s_{gk^{'}}}_{\text{intra-group interference}}\nonumber\\
&+\underbrace{\sum_{g'\ne g}\kappa d_{mk}^{-\alpha/2}\h_{mk}^{H}\B_{g'}\dd_{g'}}_{\text{inter-group interference}}
+\underbrace{\kappa d_{sk}^{-\alpha/2}h_{sk}s_{s}}_{\text{inter-tier interference}}+\underbrace{z_{k}}_{\text{noise}},
\end{align}
where $y_{mk}$ denotes the received signal of the user $k$ from the serving macro BS, $\dd_{g}$ is the sent signal from the macro BS for the users and the pico BS in group $g$, $h_{sk}\in\mathcal{CN}(0,1)$ is the channel between the pico BS and the user served by the macro BS and $s_{s}$ is the sent signal from the pico BS. Note that for the intra-group interference component, $k'$ can be $s$, i.e., the pico BS is viewed as a user served by the macro BS.

Regarding the user $k$ served by the pico BS, we have a two-hop full-duplex DF relay channel and the received signals at the pico BS and the user are given by
\begin{align}
y_{s}=&\underbrace{\kappa d_{ms}^{-\alpha/2}\h_{ms}^{H}\bb_{gs}s_{gs}}_{\text{useful signal}}+\underbrace{\sum_{k^{'}\ne s}\kappa d_{ms}^{-\alpha/2}\h_{ms}^{H}\bb_{gk^{'}}s_{gk^{'}}}_{\text{intra-group interference}}
+\underbrace{\sum_{g'\ne g}\kappa d_{ms}^{-\alpha/2}\h_{ms}^{H}\B_{g'}\dd_{g'}}_{\text{inter-group interference}}+\underbrace{z_{ms}}_{\text{noise}},\\
y_{sk}=&\underbrace{\kappa d_{sk}^{-\alpha/2}h_{sk}s_{s}}_{\text{useful signal}}+\underbrace{\sum_{g'}\kappa d_{mk}^{-\alpha/2}\h_{mk}^{H}\B_{g'}\dd_{g'}}_{\text{inter-tier interference}}+\underbrace{z_{k}}_{\text{noise}},
\end{align}
respectively. In the above equations, $y_{s}$ denotes the received signal from the macro BS at the pico BS, and $y_{sk}$ denotes the received signal of the user $k$ from the serving pico BS.

Assuming equal power allocation among the data streams, the SINR of a typical user $k$ in group $g$ served by the macro BS $\ssnr_{mk}$ can be expressed as
\begin{align}
&\ssnr_{mk}=\frac{\left|d_{mk}^{\frac{-\alpha}{2}}\h_{mk}^{H}\bb_{gk}\right|^2}{\frac{1}{\rho}+\sum\limits_{k'\ne k}\left|d_{mk}^{\frac{-\alpha}{2}}\h_{mk}^{H}\bb_{gk'}\right|^2+\sum\limits_{g'\ne g}\left\|d_{mk}^{\frac{-\alpha}{2}}\h_{mk}^{H}\B_{g'}\right\|^2+\frac{d_{sk}^{-\alpha}P_s|h_{sk}|^2}{\rho N_0}},
\end{align}
where $\rho=\frac{P_m \kappa^2}{SN_0}$ denotes the equivalent transmitted signal-to-noise ratio (SNR) at the macro BS for each data stream, $S$ is the number of data streams and $N_0$ is the noise power.

The rate of user $k$ in group $g$ served by the pico BS is decided by \cite{tse}
\begin{align}\label{eq:sinrsk}
\ssnr_{sk}&=\min\Bigg\{\frac{\frac{d_{sk}^{-\alpha}\kappa^2P_s|h_{sk}|^2}{\rho N_0}}{\frac{1}{\rho}+\sum\limits_{g'}\left |d_{mk}^{\frac{-\alpha}{2}}\h_{mk}^{H}\B_{g'}\right |^2},
\frac{\left|d_{ms}^{\frac{-\alpha}{2}}\h_{ms}^{H}\bb_{gs}\right|^2}{\frac{1}{\rho}+\sum\limits_{k'\ne s}\left|d_{ms}^{\frac{-\alpha}{2}}\h_{ms}^{H}\bb_{gk'}\right|^2+\sum\limits_{g'\ne g}\left |d_{ms}^{\frac{-\alpha}{2}}\h_{ms}^{H}\B_{g'}\right |^2}\Bigg\}\nonumber\\
& = \min\{\ssnr_{sk,g}, \ssnr_{ms}\},
\end{align}
where $\ssnr_{sk}$ represents the equivalent SINR of user $k$ served by the pico BS, $\bb_{gs}$ denotes the first-stage precoding vector at the macro BS for the pico BS, $\ssnr_{ms}$ denotes the SINR of the received signal from the macro BS at the pico BS, and $\ssnr_{sk,g}$ denotes the SINR of received signal from the serving pico BS for the user $k$ in group $g$ .

\subsection{User SINR Outage Probability}

First, we can show the following result regarding the outage probability of the user served by the macro BS.
\begin{Lem}\label{theo:userp}
The outage probability of the user served by the macro BS is given by
\begin{align}\label{eq:p1u}
P_{\tout}^{m}(x)&=1-\Pr(\ssnr_{mk}>x)\nonumber\\
&=1-\frac{\frac{ N_0\rho{\mu}_{mk,1}(x)}{xP_sd_{sk}^{-\alpha}+N_0\rho{\mu}_{mk,1}(x)}}{\prod_{i=2}^{r_g}(1-\frac{\mu_{mk,i}(x)}{\mu_{mk,1}(x)})}e^{\frac{-x}{\rho{\mu}_{mk,1}(x)}},
\end{align}
where ${\mu_{mk,i}(x),\,i=1,...,r_g}$ are the eigenvalues of $\A_{mk}(x)=d_{mk}^{-\alpha}(\A_{mk}^{'}-x\A_{mk}^{''})$ with
\begin{align}
\mathbf{A}_{mk}^{'}&=\boldsymbol{\Lambda}_g^{1/2}\U_g^{H}\bb_{gk}\bb_{gk}^{H}\U_g\boldsymbol{\Lambda}_g^{1/2},\label{eq:A1}\\
\mathbf{A}_{mk}^{''}&=\sum_{k^{'}\ne k}\boldsymbol{\Lambda}_g^{1/2}\U_g^{H}\bb_{gk^{'}}\bb_{gk^{'}}^{H}\U_g\boldsymbol{\Lambda}_g^{1/2}\notag\\
&+\sum_{g'\ne g}\boldsymbol{\Lambda}_g^{1/2}\U_g^{H}\mathbf{B}_{g'}\B_{g'}^{H}\U_g\boldsymbol{\Lambda}_g^{1/2}.\label{eq:A2}
\end{align}
Also, $\mu_{mk,1}(x)\ge \mu_{mk,2}(x)\ldots\ge\mu_{mk,r_g}(x)$. The maximum eigenvalue $\mu_{mk,1}(x)$ of $\A_{mk}(x)$ is strictly positive $\forall x\ge 0$, and the eigenvalues $\mu_{mk,2}(x),\ldots,\mu_{mk,r_g}(x)$ are non-positive $\forall x\ge 0$.
\end{Lem}
\emph{Proof:} See Appendix \ref{app:userp} for details.\hfill$\square$

\begin{Rem}
The interference from the pico BS is reflected in the numerator of (\ref{eq:p1u}). In case of no pico BS, the outage probability of the user can be expressed as
\begin{align}
P_{\tout}^{m}(x)&=1-\frac{e^{\frac{-x}{\rho{\mu}_{mk,1}(x)}}}{\prod_{i=2}^{r_g}(1-\frac{\mu_{mk,i}(x)}{\mu_{mk,1}(x)})}.
\end{align}
\end{Rem}

Regarding the user served by the pico BS, we immediately have the following result.
\begin{Lem1}
In the relay-assisted mmWave system, the outage probability of the user served by the pico BS is given by
\begin{align}
P_{\tout}^{s}(x)&=1-\Pr(\min\{\ssnr_{ms},\ssnr_{sk,g}\}>x)\nonumber\\
&=1-\Pr(\ssnr_{ms}>x)\Pr(\ssnr_{sk,g}>x).
\end{align}
\end{Lem1}

The above result is obvious since when the SINR of the link between either the macro- and pico BS link or the pico BS and user is smaller than $x$, the resultant relay channel will be in outage \cite{mmwaverelay}.

Then, we can obtain the user SINR outage probability associated with the relay, i.e., the pico BS, as follows.
\begin{Corr}
The outage probability of the user served by the pico BS is given by
\begin{align}\label{eq:p2u}
P_{\tout}^{s}(x)=1-\frac{e^{\frac{-x}{\rho{\mu}_{sk,g,1}(x)}}}{\prod_{i=2}^{r_{sk}}(1-\frac{\mu_{sk,g,i}(x)}{\mu_{sk,g,1}(x)})}
\frac{e^{\frac{-x}{{\rho\mu}_{ms,1}(x)}}}{\prod_{i=2}^{r_{g}}(1-\frac{\mu_{ms,i}(x)}{\mu_{ms,1}(x)})},
\end{align}
where ${\mu_{ms,i}(x):i=1,\ldots,r_{g}}$ and ${\mu_{sk,g,i}(x):i=1,\ldots,r_{sk}}$ are the eigenvalues of $\A_{ms}(x)=d_{ms}^{-\alpha}(\A_{ms}^{'}-x\A_{ms}^{''})$ and $\A_{sk}(x)=d_{sk}^{-\alpha}(\A_{sk}^{'}-x\A_{sk}^{''})$, respectively, and $r_{sk}$ denotes the rank of $\A_{sk}(x)$. The expressions for $\A_{ms}^{'}$ and $\A_{ms}^{''}$ are similar to the ones in (\ref{eq:A1}) and (\ref{eq:A2}), respectively, albeit $\bb_{gs}$ instead of $\bb_{gk}$, whereas
$\A_{sk}^{'}=\text{diag}(\T_{sk}^{'},\mathbf{0^{'}})$ and $\A_{sk}^{''}=\text{diag}(\0^{''},\T_{sk}^{''})$ with
\begin{align}
\T_{sk}^{'}&= \frac{\kappa^2P_s}{\rho N_0}\mathbf{1},\\
\mathbf{T}_{sk}^{''}&=\frac{d_{mk}^{-\alpha}}{d_{sk}^{-\alpha}}\sum_{g'}\tilde{\boldsymbol{\Lambda}}_g^{1/2}\tilde{\U}_g^{H}\mathbf{B}_{g'}\B_{g'}^{H}\tilde{\U}_g\tilde{\boldsymbol{\Lambda}}_g^{1/2},\label{eq:Ask1}
\end{align}
where the first element of $\mathbf{1}\in\mathbb{C}^{M\times M}$ is 1 with other elements all being zero, and the definitions of $\tilde{\U}_g$ and $\tilde{\boldsymbol{\Lambda}}_g$ are similar to (\ref{eq:hsk}).  Note that $\mathbf{\0^{'}}$ and $\mathbf{\0^{''}}$ are all $M\times M$ zero matrix.

The derivations are similar to Appendix B except for the definitions of $\A_{sk}^{'}$ and $\A_{sk}^{''}$, since considering the expression for $\ssnr_{sk,g}$ in (\ref{eq:sinrsk}), we can define
\begin{small}
\begin{align}
Z_{sk} &= d_{sk}^{-\alpha} \frac{P_s|h_{sk}|^2}{\rho N_0} - x \left(\frac{1}{\rho}+\sum\limits_{g'}\left\|d_{mk}^{\frac{-\alpha}{2}}\mathbf{w}_{mk}\boldsymbol{\Lambda}_g^{1/2}\U_g^{H}\B_{g'}\right\|^2\right)\nonumber\\
&=d_{sk}^{-\alpha}\mathbf{w}^H\A_{sk}^{'}\mathbf{w} - xd_{sk}^{-\alpha}\mathbf{w}^H\A_{sk}^{''}\mathbf{w}-\frac{x}{\rho},
\end{align}
\end{small}
where $\mathbf{w}=\left[\begin{array}{l}\w_{sk}^{'}\\ \w_{mk}\end{array}\right]$ with $\w_{sk}^{'}\in \mathbb{C}^{M\times 1}$ and the first element of $\w_{sk}^{'}$ is $h_{sk}$ while others are augmented and irrelevant independent $\mathcal{CN}(0,1)$ random variables.

Again, we assume
\begin{align}
\mu_{ms,1}(x)\ge \ldots\mu_{ms,r_{g}}(x),\\
\mu_{sk,g,1}(x)\ge \ldots\mu_{sk,g,r_{sk}}(x).
\end{align}
Still, we know that the maximum eigenvalue $\mu_{sk,g,1}(x)$ of $\A_{sk}(x)$ and $ \mu_{ms,1}(x)$ of $\A_{ms}(x)$ are strictly positive $\forall x\ge0$ and the eigenvalues $\mu_{sk,g,2}(x), \ldots, \mu_{sk,g,r_{sk}}(x)$ and $\mu_{ms,2}(x), \ldots, \mu_{ms,r_{g}}(x)$ are non-positive $\forall x\ge 0$.
\end{Corr}

\subsection{ Cell SINR Outage Probability}

Above, we have obtained the single user outage probability in the considered cellular network. Note that the cell is in outage whenever there is one user in outage, i.e., the coverage to all users cannot be guaranteed. The cell SINR outage probability can be characterized below.
\begin{Lem}\label{theo:cellp}
For the macro BS serving $K$ users employing JSDM-PGP with the aid of a single antenna pico BS, where the users are divided into $G$ groups with $\sum_{g=1}^{G}K_g=K$, the cell SINR outage probability is given by
\begin{align}\label{eq:cellout}
\hat{P}_{\sout}(x)&=\frac{\sum\limits_{i=0}^{K_g}\dbinom{K_g}{i}p_{gm}^{K_g-i}(1-p_{gm})^i\Bigg(\sum\limits_{k=1}^{K_g-i}\overline{P}_{\tout}^{m,gk}(x)+\sum\limits_{k=1}^{i}\overline{P}_{\tout}^{s,gk}(x)\Bigg)+\sum_{g'\ne g}\sum_{k}\overline{P}_{\tout}^{m,g'k}(x)}{K},
\end{align}
where $i$ is the number of users served by the pico BS that can take any value between 0 and $K_g$, $\overline{P}_{\tout}^{m,gk}(x)$, $\overline{P}_{\tout}^{s,gk}(x)$ and $\overline{P}_{\tout}^{m,g'k}(x)$ are the average outage probabilities of the user $k$ associated with the macro BS and the pico BS in the group $g$ and the user $k$ associated with the macro BS in other groups $g'$, respectively, and are given by 
\begin{small}
\begin{align}
\overline{P}_{\tout}^{m,gk}(x)&=\Upsilon-A_1(x)\Bigg (\int_{\theta_g-\Delta_g}^{\theta_g-\theta_0}\int_{0}^{R}\frac{ N_0\rho\mathbf{\Xi}_{mk,1}(x)l^{1-\alpha}}{xP_sd_{sk}^{-\alpha}+N_0\rho\mathbf{\Xi}_{mk,1}(x)l^{-\alpha}}e^{\frac{-x}{\rho\mathbf{\Xi}_{mk,1}(x)}l^{\alpha}}
\,dld\beta+\int_{\theta_g-\theta_0}^{\theta_g+\theta_0}\int_{\ell_{1}(\beta)}^{R}\nonumber\\
&\frac{ N_0\rho\mathbf{\Xi}_{mk,1}(x)l^{1-\alpha}}{xP_sd_{sk}^{-\alpha}+N_0\rho\mathbf{\Xi}_{mk,1}(x)l^{-\alpha}}
e^{\frac{-x}{\rho\mathbf{\Xi}_{mk,1}(x)}l^{\alpha}}
\,dld\beta
+\int_{\theta_g-\theta_0}^{\theta_g+\theta_0}\int_{0}^{d_{ms}}\frac{ N_0\rho\mathbf{\Xi}_{mk,1}(x)l^{1-\alpha}}{xP_sd_{sk}^{-\alpha}+N_0\rho\mathbf{\Xi}_{mk,1}(x)l^{-\alpha}}\nonumber\\
&e^{\frac{-x}{\rho\mathbf{\Xi}_{mk,1}(x)}l^{\alpha}}\,dld\beta
+\int_{\theta_g+\theta_0}^{\theta_g+\Delta_g}\int_{0}^{R}
\frac{ N_0\rho\mathbf{\Xi}_{mk,1}(x)l^{1-\alpha}}{xP_sd_{sk}^{-\alpha}+N_0\rho\mathbf{\Xi}_{mk,1}(x)l^{-\alpha}}e^{\frac{-x}{\rho\mathbf{\Xi}_{mk,1}(x)}l^{\alpha}}\,dld\beta \Bigg ),\label{eq:p1a}\\
\overline{P}_{\tout}^{s,gk}(x)&=\frac{\Delta_g}{\sum_{g'}\Delta_{g'}}-\Upsilon-A_2(x)\int_{\theta_g-\theta_{0}}^{\theta_g+\theta_{0}}\int_{d_{ms}}^{\ell_{1}(\beta)}\frac{le^{\frac{-x}{\rho d_{sk}^{-\alpha}\mathbf{\Xi}_{sk,1}(l,\beta,x)}}}{\prod_{i=2}^{r_{sk}}(1-\frac{\mathbf{\Xi}_{sk,i}(l,\beta,x)}{\mathbf{\Xi}_{sk,1}(l,\beta,x)})} \,dld\beta, \label{eq:p2a}\\
\overline{P}_{\tout}^{m,g'k}(x)&=\frac{\Delta_{g'}}{\sum_{g'}\Delta_{g'}}-A_1(x)\int_{\theta_{g'}-\Delta_{g'}}^{\theta_{g'}+\Delta_{g'}}\int_{0}^{R}\frac{ N_0\rho\mathbf{\Xi}_{mk,1}(x)l^{1-\alpha}}{xP_sd_{sk}^{-\alpha}+N_0\rho\mathbf{\Xi}_{mk,1}(x)l^{-\alpha}}e^{\frac{-x}{\rho\mathbf{\Xi}_{mk,1}(x)}l^{\alpha}}\,dld\beta, \label{eq:p3a}
\end{align}
\end{small}where $l$ is an integration variable denoting the distance between the user and the macro BS, $\beta$ is an integration variable in terms of the AOA, and 
\begin{align}
\theta_0&=2\arcsin\left(\frac{r}{2d_{ms}}\right),\\
\Upsilon&=\frac{\Delta_{g}}{\sum_{g'}\Delta_{g'}}+\frac{d_{ms}^2\theta_0}{R^2\sum_{g'}\Delta_{g'}}-\frac{1}{2R^2\sum_{g'}\Delta_{g'}}\bigg(d_{ms}^2\sin(2\theta_0)
+2r^2\theta_0\nonumber\\
&+2d_{ms}\sin(\theta_0)\sqrt{r^2-d_{ms}^2\sin^2(\theta_0)}+2r^2\arcsin(\frac{d_{ms}}{r}\sin(\theta_0))\bigg ),\label{eq:psi}\\
\ell_{1}(\beta)&=d_{ms}\cos(\beta-\theta_g)+\sqrt{r^2-d_{ms}^2\sin^2(\beta-\theta_g)},\label{eq:ell}\\
A_1(x)&=\frac{1}{R^2\sum\limits_{g'}\Delta_{g'}}\frac{1}{\prod_{i=2}^{r_g}(1-\frac{\mathbf{\Xi}_{mk,i}(x)}{\mathbf{\Xi}_{mk,1}(x)})},\label{eq:a1x}\\
A_2(x)&=\frac{e^{\frac{-x}{\rho d_{ms}^{-\alpha}\mathbf{\Xi}_{ms,1}(x)}}}{\prod_{i=2}^{r_{g}}(1-\frac{\mathbf{\Xi}_{ms,i}(x)}{\mathbf{\Xi}_{ms,1}(x)})}\frac{1}{R^2\sum\limits_{g'}\Delta_{g'}},\label{eq:a2x}\\
d_{sk}&=\sqrt{(l-d_{ms}\cos(\beta-\theta_g))^2+(d_{ms}\sin(\beta-\theta_g))^2},\label{eq:dsk}
 \end{align}
$\mathbf{\Xi}_{mk}(x)$ is the diagonal eigenvalues vector of $(\A_{mk}^{'}-x\A_{mk}^{''})$, $\mathbf{\Xi}_{ms}(x)$ is the diagonal eigenvalues vector of $(\A_{ms}^{'}-x\A_{ms}^{''})$, and $\mathbf{\Xi}_{sk}(l,\beta,x)$ is the diagonal eigenvalues matrix of $(\A_{sk}^{'}-x\A_{sk}^{''})$.
\end{Lem}
\emph{Proof:} See Appendix \ref{app:cellp} for details.\hfill$\square$

\begin{Rem}
Note that (\ref{eq:cellout}) contains three different cases for the user outage probability. Specifically, the users in group $g$ with a pico BS can be served by the macro BS or pico BS depending on the association strategy in Section \ref{sec:association}, while the users in other groups can only be served by the macro BS. The average user outage probability for the three cases are different as shown in (\ref{eq:p1a})-(\ref{eq:p3a}), respectively.
\end{Rem}

\begin{Rem}
Note that in the presence of the interference from the macro BS, the eigenvalues of $(\A_{sk}^{'}-x\A_{sk}^{''})$ is related to $d_{sk}$ as can be seen in (\ref{eq:Ask1}), which varies with the integral variables $l$ and $\beta$ from (\ref{eq:dsk}). Hence, we define $\mathbf{\Xi}_{sk,i}(l,\beta,x)$ as a function of $l$ and $\beta$ as well.
\end{Rem}

In Theorem \ref{theo:cellp}, we only consider a fixed partition of user groups that satisfy $\sum_{g=1}^{G}K_g=K$. If the $K$ users are divided into $G$ groups randomly, there will be $\frac{(K+G-1)!}{(G-1)!K!}$ possible cases according to \cite{combinatorics}. We denote the set of the partitions as $\Omega$. After characterizing the cell outage probability for each partition according to Theorem \ref{theo:cellp}, we have the following result characterizing the average cell outage probability immediately.
\begin{Lem1}\label{prop:random}
For the macro BS serving $K$ users employing JSDM-PGP with the aid of a single antenna pico BS, the average cell SINR outage probability is given by
\begin{align}
\tilde{P}_{\sout}(x)&=\sum_{\Omega}\Pi_{g'=1}^{G}p_{g'}^{K_{g'}}\hat{P}_{\sout}(x),
\end{align}
where $K_{g'}$ denotes the number of users in group $g'$ and $\Pi_{g'=1}^{G}p_{g'}^{K_{g'}}$ denotes the probability of one user group partition with $\sum_{g'=1}^{G}K_{g'}=K$ and
\begin{align}
p_{g'}&=\frac{\Delta_{g'}}{\sum_{g}\Delta_g}, g'=1,2,...,G.
\end{align}
\end{Lem1}

\begin{Exm}
 For the special case of $G=2$, $\frac{(K+G-1)!}{(G-1)!K!}=K+1$. Without loss of generality, we let $g=1$, i.e., the pico BS lies in the first group. The cell SINR outage probability is then given by
 \begin{small}
\begin{align}
\tilde{P}_{\sout}(x)&=\sum_{K_1=0}^{K}\frac{p_{1}^{K_{1}}p_{2}^{K-K_{1}}\Bigg(\sum\limits_{i=0}^{K_1}\dbinom{K_1}{i}p_{1m}^{K_1-i}(1-p_{1m})^i\Big(\sum\limits_{k=1}^{K_1-i}\overline{P}_{\tout}^{m,1k}(x)+\sum\limits_{k=1}^{i}\overline{P}_{\tout}^{s,1k}(x)\Big)+ \sum\limits_{k=1}^{K-K_{1}}\overline{P}_{\tout}^{m,2k}(x)\Bigg)}{K},
\end{align}
\end{small}
\end{Exm}

Note that it has been shown in \cite{stommWave}, \cite{gursoycoverage} that the mmWave systems are generally noise limited. For the noise limited systems, we can further simplify the expressions for $\overline{P}_{\tout}^{m,gk}(x)$, $\overline{P}_{\tout}^{s,gk}(x)$ and $\overline{P}_{\tout}^{m,g'k}(x)$.
\begin{Lem}\label{theo:noise}
In case of no interference, we have the following simplified expressions for $\overline{P}_{\tout}^{m,gk}(x)$, $\overline{P}_{\tout}^{s,gk}(x)$ and $\overline{P}_{\tout}^{m,g'k}(x)$
\begin{small}
\begin{align}
\overline{P}_{\tout}^{m,gk}(x)&= \Upsilon-\frac{1}{R^2\sum\limits_{g'}\Delta_{g'}}\Bigg (\frac{2\gamma(\frac{2}{\alpha},a_{gk}R^{\alpha})\Delta_{g}+2\theta_0\gamma(\frac{2}{\alpha},a_{gk}d_{ms}^{\alpha})}{\alpha a_{gk}^{\frac{2}{\alpha}}}
-\int_{\theta_g-\theta_0}^{\theta_g+\theta_0}\frac{\gamma(\frac{2}{\alpha},a_{gk}\ell_1^{\alpha}(\beta))}{\alpha a_{gk}^{\frac{2}{\alpha}}}\,d\beta\Bigg ),\label{aeq:p1}\\
\overline{P}_{\tout}^{s,gk}(x)&=\frac{\Delta_g}{\sum_{g'}\Delta_{g'}}-\Upsilon-\frac{1}{R^2\sum\limits_{g'}\Delta_{g'}}e^{\frac{-x}{\rho d_{ms}^{-\alpha}\bb_{gs}^{H}\mathbf{R}_g\bb_{gs}}}\int_{\theta_g-\theta_0}^{\theta_g+\theta_0}\int_{d_{ms}}^{\ell_{1}(\beta)}le^{\frac{-x N_0}{ P_sd_{sk}^{-\alpha}}}\,dld\beta,\label{aeq:p2}\\
\overline{P}_{\tout}^{m,g'k}(x)&= \frac{\Delta_{g'}}{\sum_{g}\Delta_{g}}-\frac{1}{R^2\sum\limits_{g}\Delta_{g}}\Bigg (\frac{2\gamma(\frac{2}{\alpha},a_{g'k}R^{\alpha})\Delta_{g'}}{\alpha a_{g'k}^{\frac{2}{\alpha}}} \Bigg ),\label{aeq:p3}
\end{align}
\end{small}
where $\gamma(t,v)$ is incomplete gamma function and $$a_{gk}=\frac{x}{\rho\bb_{gk}^{H}\mathbf{R}_g\bb_{gk}}.$$
 \end{Lem}
\emph{Proof:} See Appendix \ref{app:noise} for details.\hfill$\square$

\begin{Rem}
Obviously, we can find that a natural way to reduce the outage probability is to select $\bb_{gk}$ as the dominant eigenvectors of $\mathbf{R}_g$. In this way, $a_{gk}$ can be maximized and as a result $\overline{P}_{\tout}^{m,gk}(x)$, $\overline{P}_{\tout}^{s,gk}(x)$ and $\overline{P}_{\tout}^{m,g'k}(x)$ can be minimized.
\end{Rem}

\section{Numerical Result}\label{sec:numerical}

\begin{table}
\caption{System Parameters}\label{tab:para}
\begin{center}
\begin{tabular}{ | c | c | c | }
\hline
parameter & definition & value \\
 \hline
$\theta_1$, $\theta_2$ & Each group AOA &  $-20^{\circ},10^{\circ}$ \\
\hline
$\Delta_1$, $\Delta_2$ & Each group AS &  $20^{\circ}, 10^{\circ}$ \\
 \hline
$f_c$ & carrier frequency & 28GHz\\
\hline
$P_m$ &The macro BS power  & $46$ dBm\\
 \hline
$P_s$ & The pico BS power &  $28$ dBm \\
 \hline
$B$ & Bandwidth &  1 GHz\\
\hline
$\mathrm{NF}$ &Noise figure & 10 dB\\
\hline
 $\alpha$ & Path loss & 4 \\
\hline
$R$ & Macro cell radius &  $200$ m \\
\hline
$r$ & Pico cell radius &  $50$ m\\
\hline
$d_{ms}$ & Distance of macro and pico BS & $150$ m \\
\hline

\end{tabular}
\end{center}
\end{table}

\begin{figure}
    \centering
    \includegraphics[width=\figsize\textwidth]{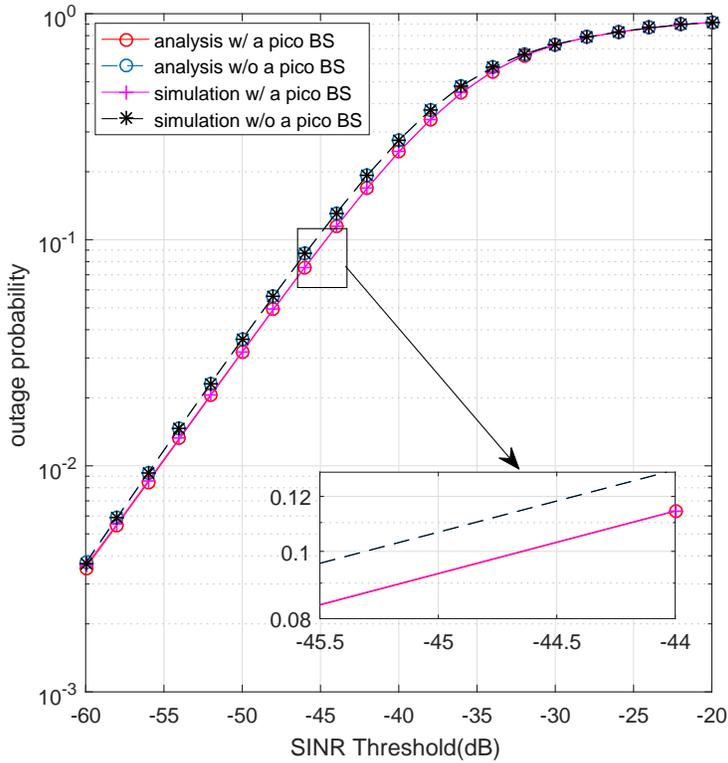}
    \caption{The cell outage probability versus SINR threshold.}
    \label{fig:result1}
\end{figure}

\begin{figure}
    \centering
    \includegraphics[width=\figsize\textwidth]{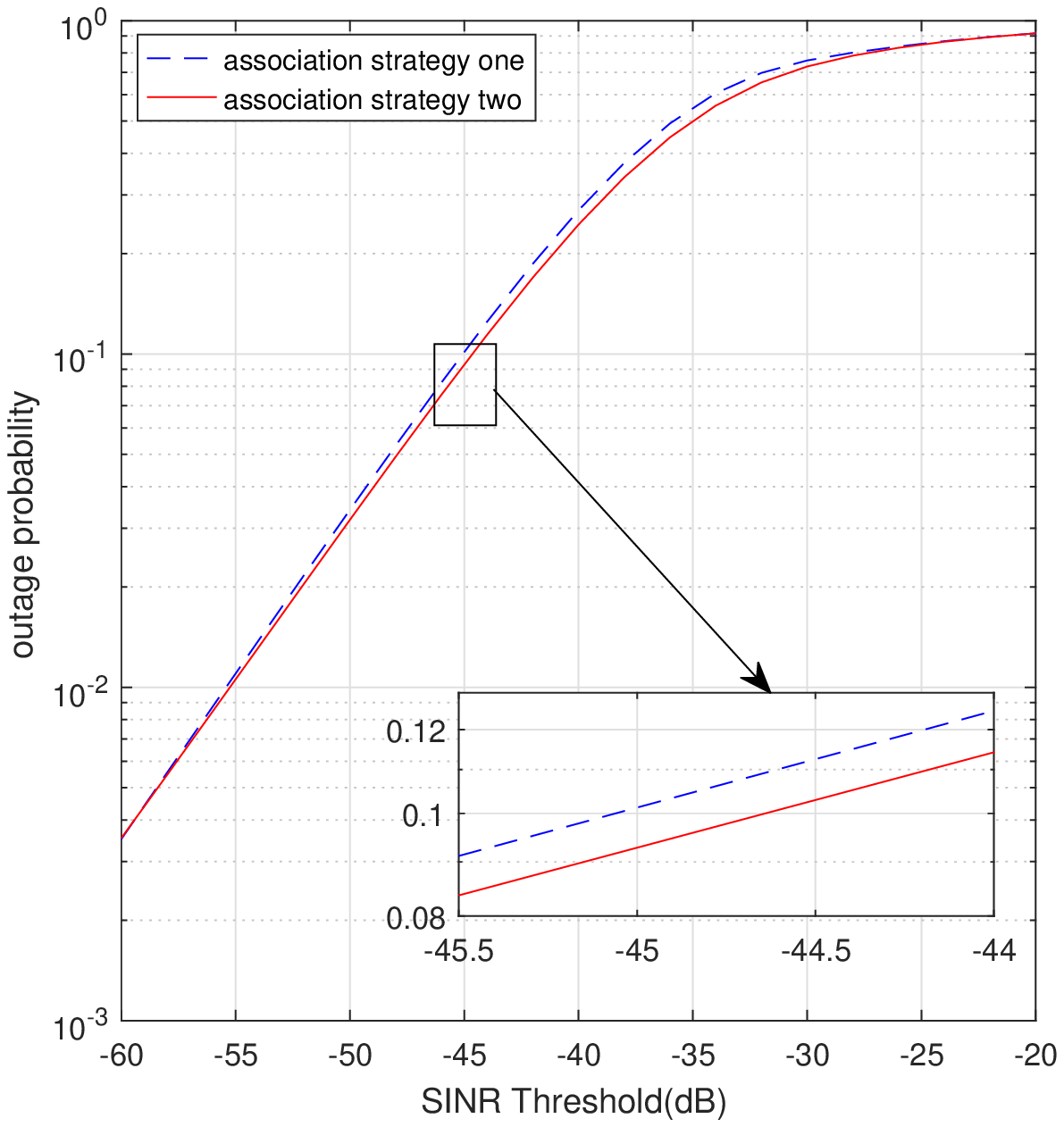}
    \caption{Different association strategies comparison .}
    \label{fig:diff-stra}
\end{figure}

\begin{figure}
    \centering
    \includegraphics[width=\figsize\textwidth]{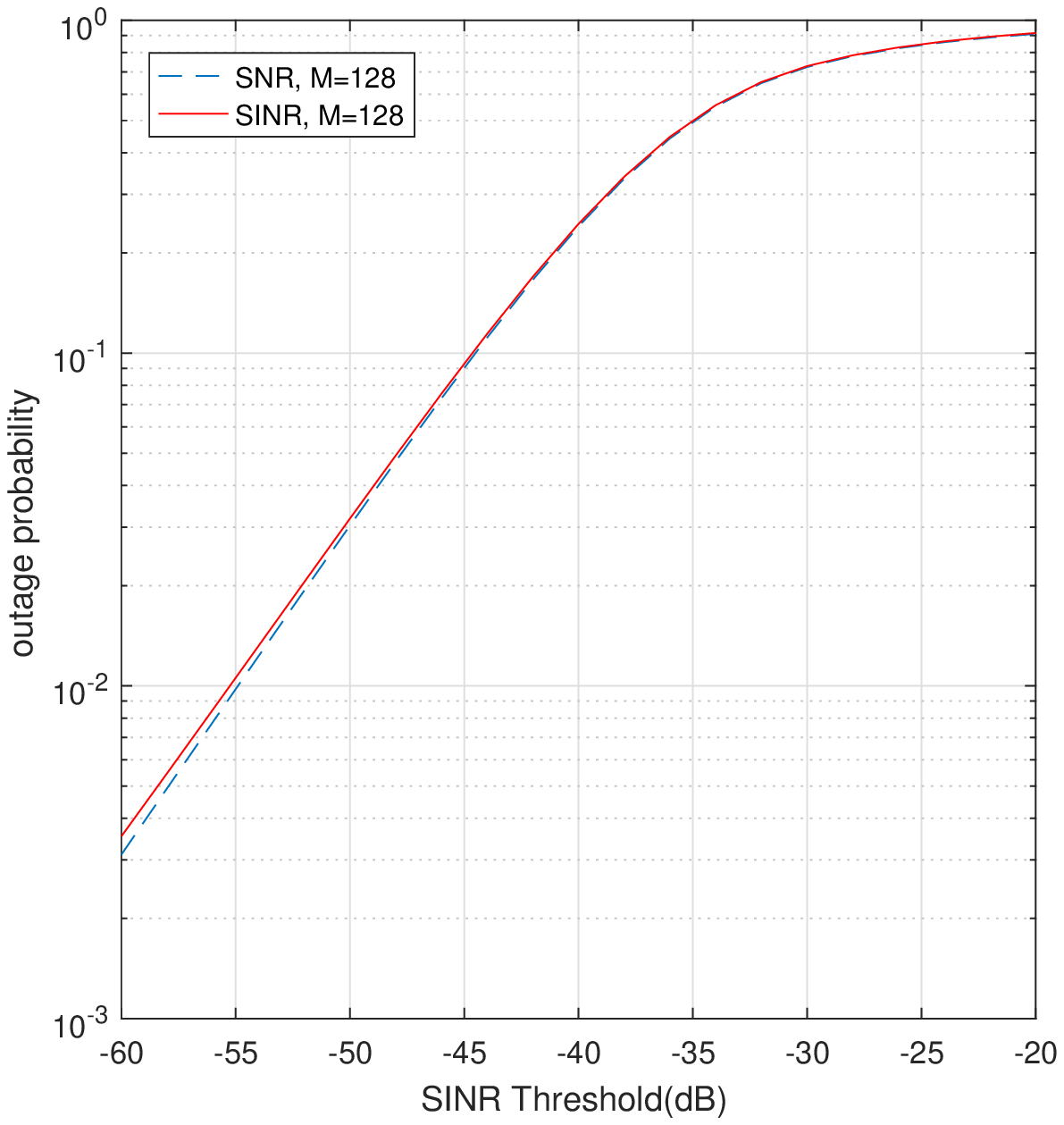}
    \caption{SNR and SINR comparison.}
    \label{fig:comp}
\end{figure}

\begin{figure}
    \centering
    \includegraphics[width=\figsize\textwidth]{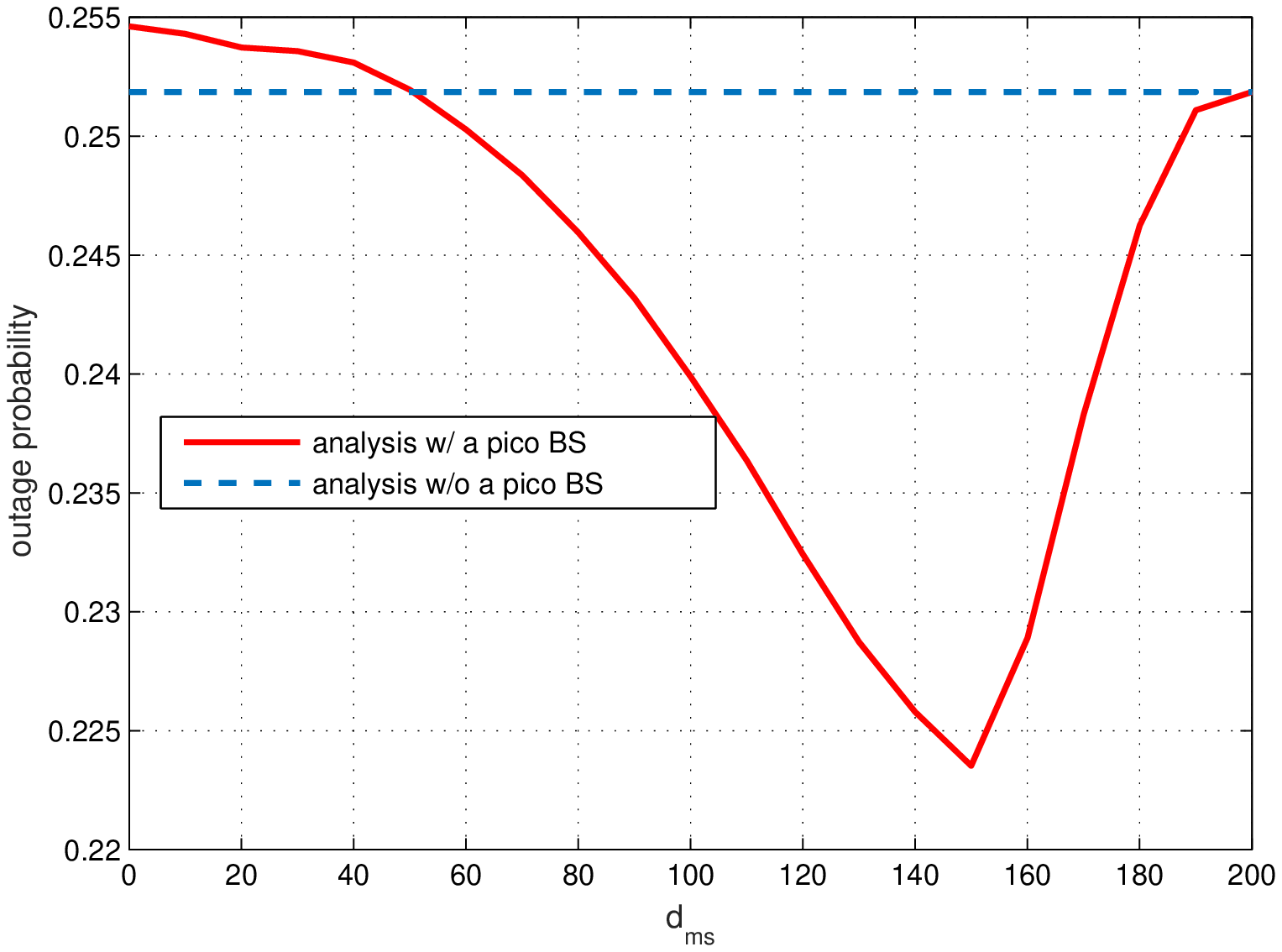}
    \caption{The cell outage probability versus $d_{ms}$.}
    \label{fig:d_ms}
\end{figure}

\begin{figure}
    \centering
    \includegraphics[width=\figsize\textwidth]{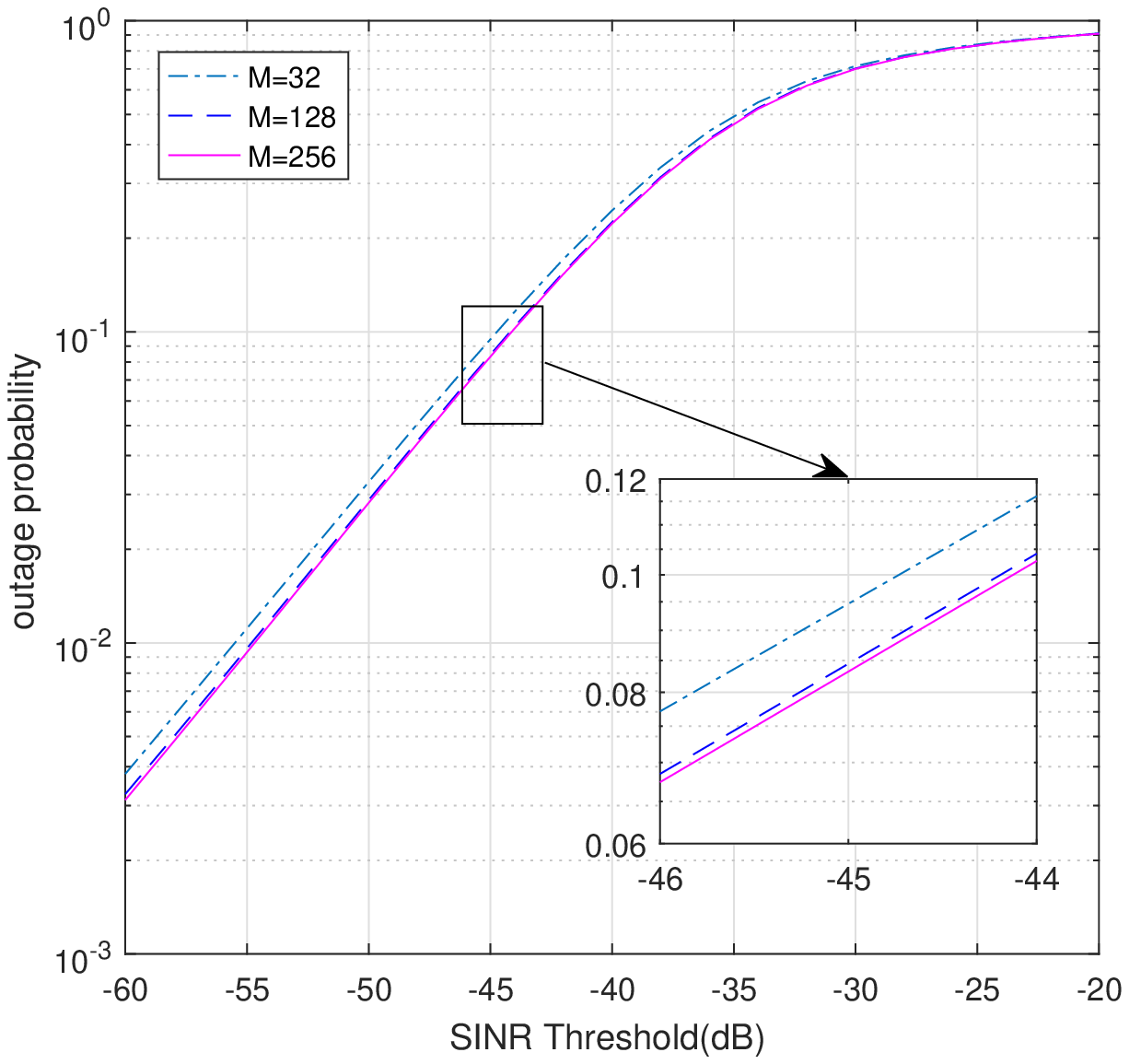}
    \caption{Different antennas comparison.}
    \label{fig:diff-ant}
\end{figure}

\begin{figure}
    \centering
    \includegraphics[width=\figsize\textwidth]{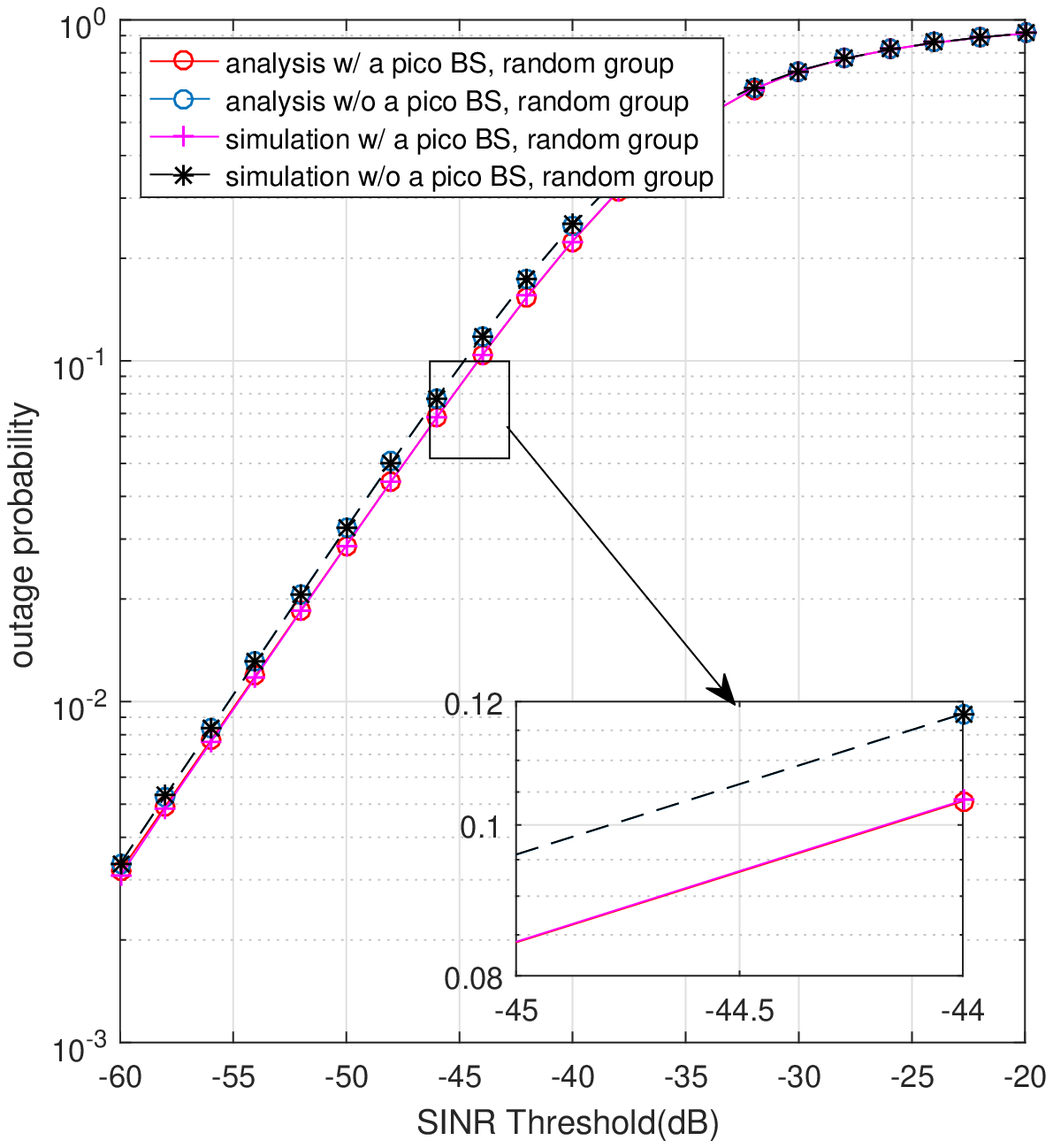}
    \caption{The cell outage probability versus SINR threshold with random group.}
    \label{fig:c-random}
\end{figure}

In this section, we evaluate the SINR outage probability of the considered relay-assisted mmWave network. We assume $K=10$ and $G=2$. The other parameters are listed in Table \ref{tab:para}. We let $g=1$, i.e., the first group contains the pico BS and $N_0(\mathrm{dBm})=-174+10\mathrm{log}_{10}(B)+\mathrm{NF(dB)}$, where $B$ and $\mathrm{NF}$ denote the bandwidth and noise figure, respectively. In Fig. \ref{fig:result1} to Fig. \ref{fig:diff-ant}, we consider a given partition of users with $K_1=7,K_2=3$. In Fig. \ref{fig:c-random}, we consider the random groups of users.

%

In Fig. \ref{fig:result1}, we plot the SINR outage probability for the user with and without a pico BS, respectively. We assume $M=128$. First, we can see that the simulation results match the analysis results, validating the theoretical analysis. Also, we can find that employing a pico BS as a relay in the mmWave cellular network employing JSDM can improve the outage performance, e.g., around 1 dB increase at $P_{\text{out}}=0.1$. Henceforth, we only show the plots of the theoretical analysis.

In Fig. \ref{fig:diff-stra}, ``association strategy one" denotes the strategy (\ref{eq:asso1}) and ``association strategy two" denotes the strategy (\ref{eq:usera}) we proposed. Comparing the two strategies, we can see that the association strategy we proposed is better and the association strategy ignoring the relay channel is almost the same as the performance as if there is no pico BS. This is generally because that certain users may experience worse performance if associated with the pico BS instead due to the relay channel formed, e.g., regions S1 and S2 in Fig. \ref{fig:2d}.

In Fig. \ref{fig:comp}, we compare the SINR outage probability with the SNR outage probability. Note that the two curves are close to each other, implying that the heterogeneous mmWave network employing JSDM is still noise-limited in accordance with the previous findings in \cite{stommWave}, \cite{gursoycoverage}, \cite{coveragandrate}. This means that noise power is still the limiting factor of the system performance for JSDM, and hence some simplified first-stage beamforming strategy can be used instead of nulling the inter-group interference considered in literature.

In Fig. \ref{fig:d_ms}, we plot the SINR outage probability as $d_{ms}$ varies. We assume $\snr=-40$ dB. We can find from the figure that the outage probability first decreases as $d_{ms}$ increases and then increases. There is an optimal value for $d_{ms}$ to achieve the smallest outage probability. This is generally because $p_{gs}$ first increases and achieves the largest value when $d_{ms}=R-r$ and then decreases. Then, the benefit introduced by the relay pico BS changes correspondingly.

In Fig. \ref{fig:diff-ant}, we compare the SINR outage probability with different number of the macro BS antennas. We can see that as the number of antennas increases, the performance improvement vanishes when $M$ is large enough, e.g., the outage probabilities for $M=128$ and $M=256$ are almost the same. In other words, increasing the macro BS antenna cannot always improve the coverage probability and after certain value, the revenue of increase in macro BS antenna can be negligible.

In Fig. \ref{fig:c-random}, we plot the SINR outage probability with random groups. Again, we can find the performance improvement due to the DF relay introduced by the pico BS.

\section{Conclusion}

In this paper, we have provided a general analytical framework to compute the SINR outage probability of a relay-assisted mmWave cellular network employing JSDM. We have assumed that the full-duplex pico BS employs DF protocol. We have analyzed the cell SINR outage probability of the considered network. Numerical evaluations in consistence with the theoretical analysis have been provided. We have shown that employing pico BS can be useful for a macro BS employing JSDM scheme. Moreover, we have found that the system is still noise-limited, which can help simplify the design of JSDM schemes.

\appendix
\subsection{Proof of Proposition \ref{prop:p1}}\label{app:p1}

\begin{figure}
    \centering
    \includegraphics[width=0.27\textwidth]{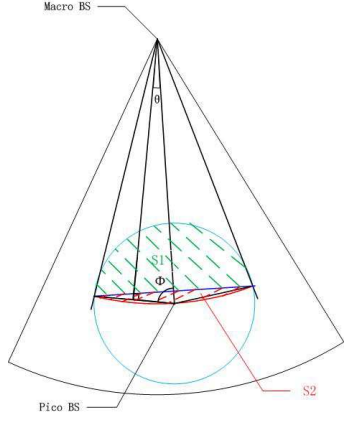}
    \caption{Illustration of the user association.}
    \label{fig:2d}
\end{figure}

First note that with the division of groups of users, we have the total area for possible user locations given by $\Delta_gR^2$. According to (\ref{eq:usera}), we must have $d_{mu}>d_{ms}$ for the user associated with the pico BS. In addition, through simple geometry computations, $\kappa^2P_sd_{su}^{-\alpha}\ge \kappa^2P_md_{mu}^{-\alpha}$ yields possible user positions in a circle. Without loss of generality, we assume that the radius of the pico BS is chosen to satisfy the previous condition. Therefore, the proposed user association strategy can be shown in Fig. \ref{fig:2d}, the region in the circle without any fill in denotes the possible locations that the user is associated with the pico BS. Then, considering the uniform distribution of the users, in the group $g$, we have
\begin{align}\label{eq:p2}
p_{gs}&=\frac{\pi r^2-S_1-S_2}{\Delta_gR^2},
\end{align}
where $S_1$ is the area of the region filled with dots S1 and $S_2$ is the area of the region filled with dashed line S2. Define $\theta =\arcsin\frac{r/2}{d_{ms}}$ and $\Phi=\frac{\pi}{2}-\theta $. After simple geometry analysis, we have
\begin{align}
S_1&=r^2\Phi-\frac{1}{2}r^2\sin2\Phi,\label{eq:s1}\\
S_2&=d_{ms}^22\theta -\frac{1}{2}d_{ms}^2\sin4\theta.\label{eq:s2}
\end{align}
Substituting (\ref{eq:s1}) and (\ref{eq:s2}) into (\ref{eq:p2}) gives us the results in (\ref{eq:p1s}) and (\ref{eq:p2s}).\hfill$\square$

\subsection{Proof of Theorem \ref{theo:userp}}\label{app:userp}

Similar to \cite{jsdmcoverage}, for a given $h_{sk}$, we define
\begin{align}
Z_{mk} &= \left|d_{mk}^{\frac{-\alpha}{2}}\h_{mk}^{H}\bb_{gk}\right|^2
- x \left(\sum\limits_{k'\ne k}\left|d_{mk}^{\frac{-\alpha}{2}}\h_{mk}^{H}\bb_{gk'}\right|^2+\sum\limits_{g'\ne g}\left\|d_{mk}^{\frac{-\alpha}{2}}\h_{mk}^{H}\B_{g'}\right\|^2\right)\nonumber\\
&-x\left(\frac{1}{\rho}+\frac{d_{sk}^{-\alpha}P_s|h_{sk}|^2}{\rho N_0}\right).
\end{align}
Then, for a given $h_{sk}$, we can obtain
\begin{align}\label{eq:z0}
\Pr(\ssnr_{mk}>x|h_{sk})=\Pr(Z_{mk}>0) = \frac{e^{\frac{-x}{\xi{\mu}_{mk,1}(x)}}}{\prod_{i=2}^{r_g}(1-\frac{\mu_{mk,i}(x)}{\mu_{mk,1}(x)})},
\end{align}
with $\xi=\frac{\rho}{1+\frac{d_{sk}^{-\alpha}P_s|h_{sk}|^2}{N_0}}$ and $\mu_{mk,i}(x)$ defined in the theorem. Since $|h_{sk}|^2$ is exponentially distributed with unit mean, we have
\begin{align}
P_{\tout}^{m}(x)&= 1-\Pr(\ssnr_{mk}>x)=1-\mathbb{E}_{h_{sk}}\{ \Pr(\ssnr_{mk}>x|h_{sk}) \}\nonumber\\
&=1-\frac{1}{\prod_{i=2}^{r_g}(1-\frac{\mu_{mk,i}(x)}{\mu_{mk,1}(x)})}e^{\frac{-x}{\rho{\mu}_{mk,1}(x)}}
\int_{0}^{\infty}e^{\frac{-xd_{sk}^{-\alpha}P_s}{N_0\rho{\mu}_{mk,1}(x)}t}e^{-t}\,dt\nonumber\\
&=1-\frac{\frac{ N_0\rho{\mu}_{mk,1}(x)}{xP_sd_{sk}^{-\alpha}+N_0\rho{\mu}_{mk,1}(x)}}{\prod_{i=2}^{r_g}(1-\frac{\mu_{mk,i}(x)}{\mu_{mk,1}(x)})}e^{\frac{-x}{\rho{\mu}_{mk,1}(x)}},
\end{align}
proving the result in the theorem.\hfill$\square$

\subsection{Proof of Theorem \ref{theo:cellp} }\label{app:cellp}

%

Note that the user can be served either directly by the macro BS or by the intermediate pico BS. Depending on the specific location of the users, we may have different number of users served by the macro BS or the pico BS.  $K_g$ is the number of the users located in group $g$ and this group contains a pico BS. For simplicity, we define events $$X=\{K_g\,\text{users in group }g\},$$ $$U=\{i\quad\text{users associated with the pico BS in group }g\},$$ $$V=\{\text{user}\, k\, \text{is served by the macro BS in group }g\},$$ $$W=\{\text{user}\, k\, \text{is served by the pico BS in group }g\},$$ and $$Y=\{\text{user}\, k\, \text{is served by the macro BS in group }g'\}.$$ Given the fact that whenever there is one user in outage, the cell will be in outage. We can show
\begin{small}
\begin{align}
\Pr\{\text{cell outage}\} &=\Pr\{X\}\Pr\{\text{group }g \text{ is in outage}|X\}+\Pr\{X\}\Pr\{\text{other groups is in outage}|X\}\label{eq:proof0}\\
&=\frac{1}{K}\bigg(\sum_{i=0}^{K_g}\Pr\{U\}\bigg(\sum_{k=1}^{K_g-i}\E\{\Pr\{\snr_{k}<x|U,V,X\}\Pr\{V|U,X \}\} \nonumber\\ %
&+ \sum_{k=1}^{i}\E\{\Pr\{\snr_{k}<x|U, W,X\}\Pr\{ W| U,X\}\} \bigg)
+\sum_{g'\neq g}\sum_{k=1}^{K_{g'}}\E\{\Pr\{\snr_{k}<x|X\}\}\bigg),\label{eq:proof1}\\
&= \frac{1}{K}\Bigg(\sum_{i=0}^{K_g}\dbinom{K_g}{i}p_{gm}^{K_g-i}(1-p_{gm})^i  \bigg(\sum_{k=1}^{K_g-i}\E_{(\beta,l)\in V}\{\Pr\{\snr_{mk}<x\}\} \nonumber\\
&+ \sum_{k=1}^{i}\E_{(\beta,l)\in W}\{\Pr\{\snr_{sk}<x\}\} \bigg)+\sum_{g'\neq g}\sum_{k=1}^{K_{g'}}\E_{(\beta,l)\in Y}\{\Pr\{\snr_{mk}<x\}\}\Bigg)\label{eq:proof3}\\
&= \frac{1}{K}\sum_{i=0}^{K_g}\dbinom{K_g}{i}p_{gm}^{K_g-i}(1-p_{gm})^i \bigg(\sum_{k=1}^{K_g-i}\E_{(\beta,l)\in V}\{P_{\tout}^{m}(x)\}+ \sum_{k=1}^{i}\E_{(\beta,l)\in W}\{P_{\tout}^{s}(x)\} \bigg) \nonumber\\
&+\frac{1}{K}\sum_{g'\neq g}\sum_{k=1}^{K_{g'}}\E_{(\beta,l)\in Y}\{P_{\tout}^{m}(x)\},\label{eq:proof5}
\end{align}
\end{small}
where the expectations in (\ref{eq:proof3}) are taken over the user distributions with $\beta$ and $l$ representing the angle and the distance between the user and the macro BS, respectively, and $\mathbb{E}_{\star}$ denotes the expectation taken over region $\star$. Note that given a partition of user groups, $\Pr\{X\}=\Pr\{X^c\}=1$ is used in (\ref{eq:proof1}). Since the outage probabilities in (\ref{eq:proof1}) are defined for one user only, $\frac{1}{K}$ is multiplied to reflect the average cell outage probability considering the cumulative distribution of the SINR of $K$ users. (\ref{eq:proof3}) comes from the fact that $i$ can take any value between 0 and $K_g$ and for given $i$, $\Pr\{U \}=\dbinom{K_g}{i}p_{gm}^{K_g-i}(1-p_{gm})^i$ with $p_{gm}$ defined in (\ref{eq:p1s}) is incorporated. When user $k$ is served by the macro or pico BS, the received SINR is given by $\snr_{mk}$ or $\snr_{sk}$, respectively, for (\ref{eq:proof3}).

Now, for a given user associated with the macro BS or the pico BS, we can obtain the outage probability from (\ref{eq:p1u}) or (\ref{eq:p2u}) accordingly. Then, we can take the expectation of the resultant outage probability over the user locations associated with the macro BS or the pico BS, respectively. Therefore, we need to characterize the region that the user is associated with the pico BS. Define $\theta_0=2\theta$. We first have the following result.
\begin{Lem1}\label{prop:polar}
Let the macro BS location be the origin of the polar coordinates. Consider the region division of a specific group, the region in which the user is served by the pico BS is described by the polar coordinate as
\begin{align}
d_{ms}\le l\le \ell_1(\beta), \,\,\beta\in(\theta_g-\theta_0,\theta_g+\theta_0),
\end{align}
where $\ell_1(\beta)=d_{ms}\cos(\beta-\theta_g)+\sqrt{r^2-d_{ms}^2\sin^2(\beta-\theta_g)}.$
\end{Lem1}
\emph{Proof:} First, we know $d_{mk}>d_{ms}$ from (\ref{eq:usera}), i.e., $l\ge d_{ms}$. Then, if we view the line connecting the macro BS and the pico BS as the $x$-axis and express the planar coordinates of the locations on the edge of the pico-cell, we have
\begin{align}
(x-d_{ms})^2+y^2 = r^2.
\end{align}
Through the change of polar coordinate $(\beta,\ell_1(\beta))$, we can rewrite the above equation as
\begin{align}
(\ell_1(\beta)\cos(\beta-\theta_g)-d_{ms})^2+(\ell_1(\beta)\sin(\beta-\theta_g))^2 = r^2,
\end{align}
which after simple computation gives us the upperbound on $\rho$, i.e., $l\le \ell_1(\beta)$. Together, we prove the result.\hfill$\square$

After characterizing the region in which the users are served by the pico BS, we first consider the user served by the macro BS and derive the expected SINR outage probability by taking the expectation of (\ref{eq:p1u}) over the region in which that the user is associated with the macro BS. Specifically, as can be seen in Appendix \ref{app:p1}, the region is irregular and hence the expectation is generally complicated. We have
\begin{align}\label{eq:p1aexp}
\overline{P}_{\tout}^{m,gk}(x)&=\E_{k\in V}\{P_{\tout}^{m}(x)\}=\int_{\theta_g-\Delta_g}^{\theta_g-\theta_0}\int_{0}^{R}P_{\tout}^{m}f_{\beta}f_l\,dld\beta\notag\\
&+\int_{\theta_g-\theta_0}^{\theta_g+\theta_0}\int_{\ell_1(\beta)}^{R}P_{\tout}^{m}f_{\beta}f_l\,dld\beta\nonumber\\
&+\int_{\theta_g-\theta_0}^{\theta_g+\theta_0}\int_{0}^{d_{ms}}P_{\tout}^{m}f_{\beta}f_l\,dld\beta+\int_{\theta_g+\theta_0}^{\theta_g+\Delta_g}\int_{0}^{R}P_{\tout}^{m}f_{\beta}f_l\,dld\beta,
\end{align}
where $f_{\beta}=\frac{1}{2\sum_{g'}\Delta_{g'}}$ and $f_l=\frac{2l}{R^2}$ represent the uniform distribution of the users in terms of polar coordinates. Substituting (\ref{eq:p1u}) into (\ref{eq:p1aexp}), we can arrive at (\ref{eq:p1a}), where $\Upsilon$ is given by
\begin{align}
\Upsilon &=\int_{\theta_g-\Delta_g}^{\theta_g-\theta_0}\int_{0}^{R}f_{\beta}f_l\,dld\beta
+\int_{\theta_g-\theta_0}^{\theta_g+\theta_0}\int_{\ell_1(\beta)}^{R}f_{\beta}f_l\,dld\beta\nonumber\\
&+\int_{\theta_g-\theta_0}^{\theta_g+\theta_0}\int_{0}^{d_{ms}}f_{\beta}f_l\,dld\beta+\int_{\theta_g+\theta_0}^{\theta_g+\Delta_g}\int_{0}^{R}f_{\beta}f_l\,dld\beta,\\
&=\frac{\Delta_{g}}{\sum_{g'}\Delta_{g'}}+\frac{d_{ms}^2\theta_0}{R^2\sum_{g'}\Delta_{g'}}-\frac{1}{2R^2\sum_{g'}\Delta_{g'}}\Bigg( \int_{\theta_g-\theta_0}^{\theta_g+\theta_0}d_{ms}^2\cos^2(\beta-\theta_g)\,d\beta\nonumber\\
&+\int_{\theta_g-\theta_0}^{\theta_g+\theta_0}(r^2-d_{ms}^2\sin^2(\beta-\theta_g))\,d\beta
+\int_{\theta_g-\theta_0}^{\theta_g+\theta_0}2d_{ms}\cos(\beta-\theta_g)\sqrt{r^2-d_{ms}^2\sin^2(\beta-\theta_g)}\,d\beta \Bigg),\\
&=\frac{\Delta_{g}}{\sum_{g'}\Delta_{g'}}+\frac{d_{ms}^2\theta_0}{R^2\sum_{g'}\Delta_{g'}}-\frac{1}{2R^2\sum_{g'}\Delta_{g'}}\Bigg(d_{ms}^2\sin(2\theta_0)
+2r^2\theta_0\nonumber\\
&+\int_{\theta_g-\theta_0}^{\theta_g+\theta_0}2d_{ms}\cos(\beta-\theta_g)\sqrt{r^2-d_{ms}^2\sin^2(\beta-\theta_g)}\,d\beta,\Bigg)\label{eq:psi0}
\end{align}
After integral variable change, we have
\begin{align}
&\int_{\theta_g-\theta_0}^{\theta_g+\theta_0}2d_{ms}\cos(\beta-\theta_g)\sqrt{r^2-d_{ms}^2\sin^2(\beta-\theta_g)}\,d\beta \nonumber\\
&\overset{t=\sin(\beta-\theta_g)}{=}\int_{-\sin\theta_0}^{\sin\theta_0}2d_{ms}\sqrt{r^2-d_{ms}^2t^2}\,dt,\nonumber\\
&=2d_{ms}\sin(\theta_0)\sqrt{r^2-d_{ms}^2\sin^2(\theta_0)}+2r^2\arcsin(\frac{d_{ms}}{r}\sin(\theta_0)).\label{eq:integralchange}
\end{align}
where $\int\sqrt{a+bx+cx^2}dx=\frac{(2cx+b)\sqrt{a+bx+cx^2}}{4c}+\frac{\Delta}{8c}\frac{-1}{\sqrt{-c}}\arcsin(\frac{2cx+b}{\sqrt{-\Delta}}), c<0, \Delta=4ac-b^2$ is incorporated \cite[2.262.1]{mathbook}. Combining (\ref{eq:integralchange}) with (\ref{eq:psi0}) yields (\ref{eq:psi}).

Similarly, considering the region in which the user is associated with the pico BS, we have
\begin{align}
\overline{P}_{\tout}^{s,gk}(x)&=\E_{k\in W}\{P_{\tout}^{s}(x)\}=\int_{\theta_g-\theta_0}^{\theta_g+\theta_0}\int_{d_{ms}}^{\ell_{1}(\beta)}P_{\tout}^{s}(x)f_{\beta}f_l\,dld\beta.
\label{aeq:psgk}
\end{align}
Combining (\ref{eq:p2u}) with (\ref{aeq:psgk}) and after simple computations, we obtain (\ref{eq:p2a}). Note that the distance between the user and pico BS can be derived following the idea of coordinate change in Proposition \ref{prop:polar} by viewing the line connecting the macro BS and user as the $x$-axis and is given by
\begin{align}
d_{sk}=\sqrt{(l-d_{ms}\cos(\beta-\theta_g))^2+(d_{ms}\sin(\beta-\theta_g))^2}.
\end{align}

In addition, for the user group without pico BS, we have
\begin{align}\label{eq:p1aexp1}
\overline{P}_{\tout}^{m,g'k}(x)=\E_{k\in Y}\{P_{\tout}^{m}(x)\}=\int_{\theta_{g'}-\Delta_{g'}}^{\theta_{g'}+\Delta_{g'}}\int_{0}^{R}P_{\tout}^{m}f_{\beta}f_l\,dld\beta
\end{align}
Inserting (\ref{eq:p1u}) into above equation and after simple computations, we get (\ref{eq:p3a}).
\hfill$\square$

\subsection{Proof of Theorem \ref{theo:noise}}\label{app:noise}
If there is no interference, $\A_{mk}=d_{mk}^{-\alpha}\A_{mk}^{'}$ in (\ref{eq:A1}) has rank $1$. Then, we know that $\mathbf{\Xi}_{mk,i}(x)=0, i\neq1$ and
\begin{align}
\mathbf{\Xi}_{mk,1}(x)&=\text{trace}\{\A^{'}_{mk}\}= \text{trace}\{\boldsymbol{\Lambda}_g^{1/2}\U_g^{H}\bb_{gk}\bb_{gk}^{H}\U_g\boldsymbol{\Lambda}_g^{1/2}\}\nonumber\\
&=\text{trace}\{\bb_{gk}^{H}\U_g\boldsymbol{\Lambda}_g\U_g^{H}\bb_{gk}\}\label{eq:zeroint-proof0}\\
&=\bb_{gk}^{H}\mathbf{R}_g\bb_{gk}\label{eq:zeroint-proof1}
\end{align}
where $\text{trace}\{AB\}=\text{trace}\{BA\}$ is used in (\ref{eq:zeroint-proof0}) and (\ref{eq:zeroint-proof1}) makes use of the facts that the trace of a scalar is itself and $\mathbf{R}_g=\U_g\boldsymbol{\Lambda}_g\U_g^{H}$.

Similarly, we can show that $\mathbf{\Xi}_{ms,i}(x)=0,i\neq1$, and
\begin{align}
\mathbf{\Xi}_{ms,1}(x)=\bb_{gs}^{H}\mathbf{R}_g\bb_{gs},
\end{align}
and $\mathbf{\Xi}_{sk,i}(l,\beta,x)=0,i\neq1 $, and
\begin{align}
\mathbf{\Xi}_{sk,1}(l,\beta,x) = \frac{P_s}{\rho N_0}.
\end{align}

Then, we have the simplified expressions for (\ref{eq:a1x}) and (\ref{eq:a2x}) as
\begin{align}
A_1(x)&=\frac{1}{R^2\sum\limits_{g'}\Delta_{g'}},\label{eq:a1x0}\\
A_2(x)&=\frac{1}{R^2\sum\limits_{g'}\Delta_{g'}}e^{\frac{-x}{\rho d_{ms}^{-\alpha}\bb_{gs}^{H}\mathbf{R}_g\bb_{gs}}},
\end{align}

Now, the outage probability of the user served by the macro BS in (\ref{eq:p1u}) can be rewritten as
\begin{align}
P_{\tout}^{m}(x)&=1-e^{\frac{-x}{\rho{\mu}_{mk,1}(x)}}=1-e^{\frac{-x}{\rho d_{mk}^{-\alpha}\mathbf{\Xi}_{mk,1}(x)}}
\end{align}
As a result, $\overline{P}_{\tout}^{m,gk}(x)$ defined in (\ref{eq:p1a}) can be simplified to
\begin{align}
\overline{P}_{\tout}^{m,gk}(x)&= \Upsilon-A_1(x)\Bigg (\int_{\theta_g-\Delta_g}^{\theta_g-\theta_0}\int_{0}^{R}le^{\frac{-x}{\rho\mathbf{\Xi}_{mk,1}(x)}l^{\alpha}}
\,dld\beta\nonumber\\
&\hspace{-1.6cm}+\int_{\theta_g-\theta_0}^{\theta_g+\theta_0}\int_{\ell_1(\beta)}^{R}le^{\frac{-x}{\rho\mathbf{\Xi}_{mk,1}(x)}l^{\alpha}}
\,dld\beta+\int_{\theta_g-\theta_0}^{\theta_g+\theta_0}\int_{0}^{d_{ms}}le^{\frac{-x}{\rho\mathbf{\Xi}_{mk,1}(x)}l^{\alpha}}\,dld\beta\nonumber\\
&\hspace{-1.6cm}+\int_{\theta_g+\theta_0}^{\theta_g+\Delta_g}\int_{0}^{R}le^{\frac{-x}{\rho\mathbf{\Xi}_{mk,1}(x)}l^{\alpha}}\,dld\beta
 \Bigg ),
\end{align}

Substituting (\ref{eq:zeroint-proof1}) and (\ref{eq:a1x0}) into above equation, we can obtain (\ref{aeq:p1}), where $\int_{0}^{u}xe^{-ax^n}=\frac{\gamma(\frac{2}{n},au^n)}{na^{2/n}}$ is incorporated \cite[3.381.8]{mathbook}.
Similarly, we can derive the simplified expressions for $\overline{P}_{\tout}^{s,gk}(x)$ and $\overline{P}_{\tout}^{m,g'k}(x)$. \hfill$\square$

%
%
 

\begin{thebibliography}{1}

\bibitem{tom-noncooperative} T. Marzetta, ``Noncooperative cellular wireless with unlimited numbers of base station antennas, ''\emph{IEEE Trans. Wireless Commun.}, vol. 9, no. 11, pp. 3590 - 3600, Nov. 2010.
\bibitem{massiveMIMO} S. Haghighatshoar, G. Caire, ``Massive MIMO pilot decontamination and channel interpolation via wideband sparse channel estimation, '' \emph{IEEE Trans. Wireless Commun.}, vol. 16, no. 12, pp. 8316 - 8332, Dec. 2017.

\bibitem{tsrapp} T. S. Rappaport \emph{et al.}, ``Millimeter wave mobile communications for 5G cellular: it will work!'' \emph{IEEE Access}, vol. 1, pp. 335 - 349, May 2013.

\bibitem{2} W. Roh \emph{et al.}, ``Millimeter-wave beamforming as an enabling technology for 5G cellular communications: Theoretical feasibility and prototype results, '' \emph{IEEE Commun. Mag.}, vol. 52, no.2, pp. 106 - 113, Feb. 2014.


\bibitem{1} J. G. Andrews \emph{et al.}, ``What will 5G be? '' \emph{IEEE J.Sel. Areas Commun.}, vol. 32, no. 6, pp. 1065 - 1082, Jun. 2014.

\bibitem{udn} X. Ge \emph{et al.}, ``5G ultra-dense cellular networks,'' \emph{IEEE Wireless Commun.}, vol. 23, no. 1, pp. 72 - 79, Feb. 2016.

\bibitem{mmwavetutorial} J. G. Andrews \emph{et al.}, ``Modeling and analyzing millimeter wave cellular systems, '' \emph{IEEE Trans. Commun.}, vol. 65, no. 1, pp. 403 - 430, Jan. 2017.



\bibitem{coverage} T. Bai and R. W. Heath, ``Coverage and rate analysis for millimeter-wave cellular networks, '' \emph{IEEE Trans. Wireless Commun.}, vol. 14, no. 2, pp. 1100 - 1114, Feb. 2015.
\bibitem{stommWave} M. Di Renzo, ``Stochastic geometry modeling and analysis of multi-tier millimeter wave cellular networks,'' \emph{IEEE Trans. Wireless Commun.}, vol. 14, no. 9, pp. 5038 - 5057, Sep. 2015.


\bibitem{gursoycoverage} E. Turgut and M. C. Gursoy, ``Coverage in heterogeneous downlink millimeter wave cellular network, '' \emph{IEEE Trans. Commun.}, vol. 65, no. 10, pp. 4463 - 4477, Oct. 2017.

\bibitem{wcsp} G. Yao, N. Liu, Z. Pan, and X. You, ``Coverage and rate analysis for non-uniform millimeter-wave heterogeneous cellular network, '' in the 2016 International Conference on Wireless Communications \& Signal Processing (WCSP), Yangzhou, China, Oct. 2016.

\bibitem{mmwaverelay} S. Biswas, S. Vuppala, J. Xue, and T. Ratnarajah, ``On the performance of relay aided millimeter wave networks, '' \emph{IEEE J. Sel. Topics Sig. Proc.}, vol. 10, no. 3, pp. 576 - 588, Apr. 2016.



\bibitem{mmwaveantenna} X. Yu, J. Zhang, M. Haenggi, and K. B. Letaief, ``Coverage analysis for millimeter wave networks: the impact of directional antenna arrays, '' \emph{IEEE J. Sel. Areas Commun.}, vol. 35, no. 7, pp. 1498 - 1512, July 2017.

\bibitem{BScooperation} D. Maamari, N. Devroye, and D. Tuninetti, ``Coverage in mmWave cellular networks with base station co-operation, '' \emph{IEEE Trans. Wireless Commun.}, vol. 15, no. 4, pp. 2981 - 2994, Apr. 2016.


\bibitem{jsdmmmwave} A. Adhikary, E. Al Safadi, M. K. Samimi, R. Wang, G. Caire, T. S. Rappaport and A. F. Molisch, ``Joint spatial division and multiplexing for mm-Wave channels, '' \emph{IEEE J. Sel. Areas in Commun}, vol. 32, pp. 1239 - 1255, May 2014.
\bibitem{jsdm} A. Adhikary, J. Nam, J.-Y. Ahn and G. Caire,``Joint spatial division and multiplexing〞the large-scale array regime, '' \emph{IEEE Trans. Inf. Theory}, vol. 59, no. 10, pp. 6441 - 6463, Oct. 2013.



\bibitem{jsdmcoverage} J. Nam, A. Adhikary, J.-Y. Ahn and G. Caire, ``Joint spatial division and multiplexing: Opportunistic beamforming, user grouping and simplified downlink scheduling, '' \emph{IEEE J. of Sel. Topics in Sig. Proc.}, vol. 8, no. 5, pp. 876 - 890,  Oct. 2014.





\bibitem{dfbrs} K. Belbase, Z. Zhang, H. Jiang, and C. Tellambura, ``Coverage analysis of millimeter wave decode-and-forward networks with best relay selection,'' \emph{IEEE Access}, pp. 1-1, 2018.


\bibitem{coveragandrate} M. N. Kulkarni, S. Singh, and J. G. Andrews, ``Coverage and rate trends in dense urban mmWave cellular networks,'' in \emph{Proc. IEEE Global Commun. Conf. (GLOBECOM)}, Dec. 2014, pp. 3809 - 3814.


\bibitem{fd-2} D. Bharadia, E. McMilin, and S. Katti, ``Full duplex radio,'' in Proc. ACM SIGCOMM conf. Appl. Technol., Archit., Protocols Comput. Commun., Hong Kong, Aug. 2013.

\bibitem{fd-3} Z. Zhang, X. Chai, K. Long, A. V. Vasilakos, and L. Hanzo, ``Full duplex techniques for 5G networks: Self-interference cancellation, protocol design, and relay selection,'' \emph{IEEE Commun. Mag.}, vol. 53, no. 5, pp. 128 - 137, May 2015.


\bibitem{tse} J. N. Laneman, D. N. C. Tse and G. W. Wornell, ``Cooperative diversity in wireless networks: efficient protocols and outage behavior, '' \emph{IEEE Trans. Inf. Theory}, vol. 50, no. 12, pp. 3062 - 3080, Dec. 2004.

\bibitem{combinatorics} Richard A. Brualdi, \emph{Introductory  Combinatorics}, 5th ed. Pearson Press, 2009.

\bibitem{mathbook} I. S. Gradstein and I. M. Ryzhik,  \emph{Tables of Integrals, Sums, Series, and Products}, Academic Press, 2000.

\bibitem{icc18} J. Chen and D. Qiao, ``Outage Analysis of Heterogeneous mmWave Cellular Systems Employing JSDM,'' in IEEE ICC 18, Kansas City, MO, May 2018.
\end{thebibliography}
\end{document}